\documentclass[twocolumn, 11pt, times]{article}
\usepackage{graphicx}     
\usepackage{caption}     
\usepackage{amsmath}
\usepackage{lastpage}
\usepackage{fancyhdr}
\makeatletter
\renewcommand{\section}{\@startsection
{section}
{1}
{0pt}
{\baselineskip}
{0.5\baselineskip}
{\bfseries\centering}}
\makeatother

\makeatletter
\renewcommand{\subsection}{\@startsection
{subsection}
{1}
{0pt}
{\baselineskip}
{0.5\baselineskip}
{\bfseries\centering}}
\makeatother

\begin{document}

\title{\bfseries\fontsize{12}{16}\selectfont
On Scalability of Wireless Networks: A Practical Primer for Large Scale Cooperation\thanks{\fontsize{9}{11}\selectfont This work is sponsored by
the United States Air Force under Air Force Contract FA8721-05-C-0002. 
Opinions, interpretations, recommendations, and conclusions are those of the
authors and are not necessarily endorsed by the United States Government. Specifically, this work
was sponsored by Information Systems of ASD(R\&E).  Linda M. Zeger was
at MIT Lincoln Lab when this research was performed.}}

\author{Linda M. Zeger and Muriel M\'{e}dard \\ Auroral LLC and MIT RLE}

\pagestyle{fancyplain}

\maketitle

\section*{Abstract} {\itshape } An intuitive overview of the
scalability of a variety of types of wireless networks is presented.
Simple heuristic arguments are demonstrated here for scaling laws presented in other works,
as well as for conditions not previously considered  in the literature.
Unicast and multicast messages, 
topology, hierarchy, and effects of reliability protocols are discussed. We show how two
key factors, bottlenecks and erasures, can often dominate the network
scaling behavior.  Scaling of  throughput or delay with the number
of transmitting nodes, the number of receiving nodes, and the file
size is described.

\section{Introduction}  \label{intro}
The dependence of network performance on the number of nodes in the network is an emerging area of interest, as there is an increasing desire to expand  wireless  networks to larger sizes.  Rendering the scalability of networks more difficult is the widespread  advent of imagery and streaming video, which means the users must transmit or receive larger volumes of data.

The question of scalability was first considered over ten years ago in the context of
sensor networks.
The question was raised as to whether performance could be improved by increasing the density of the sensors or motes.
 Addressing this question, the work \cite{GuptaKumar} showed how the capacity scales as the number of nodes 
is increased within 
a fixed area two dimensional area.  Later works \cite{hier_Ozgur1} consider fixed node density {\it or} fixed network area.


Most of the previous works in the literature, while yielding precise results given their assumptions, do not include direct consideration of two crucial factors that in practice often dominate the network scaling laws:  bottlenecks and erasures.  In this article, we show how bottlenecks and erasures can dominate the scaling behavior in many networks, and thereby yield {\it different} scaling laws than those derived previously.  We also discuss potential mitigation techniques for bottlenecks and erasures.

We provide an overview of representative
works in the scaling literature  by using intuitive arguments to explain the mathematical results. 
Our goal is to provide a didactic perspective on scaling which highlights the issues of its practical application.

In Section~\ref{bottleneck}, scaling with a range of bottleneck types is illustrated.  Potential methods to circumvent bottlenecks are discussed in Section~\ref{no_bottleneck}.  The impact on  scaling of protocols designed to handle erasures   is discussed in Section~\ref{erasures}.  

\section{Bottlenecks} \label{bottleneck}
Some previous scaling results, as well as new results presented here, can be viewed as a 
consequence of
simple bottlenecks.  Such bottlenecks can arise at most nodes as in Figure~\ref{unicast_fig}a, a group of nodes as in Figure~\ref{unicast_fig}b, 
or at designated cluster heads as in Figure~\ref{unicast_fig}c.
Scaling laws for each
of these types of bottlenecks are found by determining how many flows
must share a bottleneck node.



\subsection{When Every Node is the Bottleneck: Point-to-Point Transmissions} \label{fixed}
In this section, multiple unicast transmissions in a multi-hop network are considered, when
 there are no  packet losses nor mobility, as in \cite{GuptaKumar}.
We use a simple heuristic argument to illustrate the results of
\cite{GuptaKumar}, and then extend it to three dimensions.  

\begin{figure} [h!]
\centering
{\includegraphics[width=2in,keepaspectratio=1,
viewport=2in 1.1in 5in 9in]{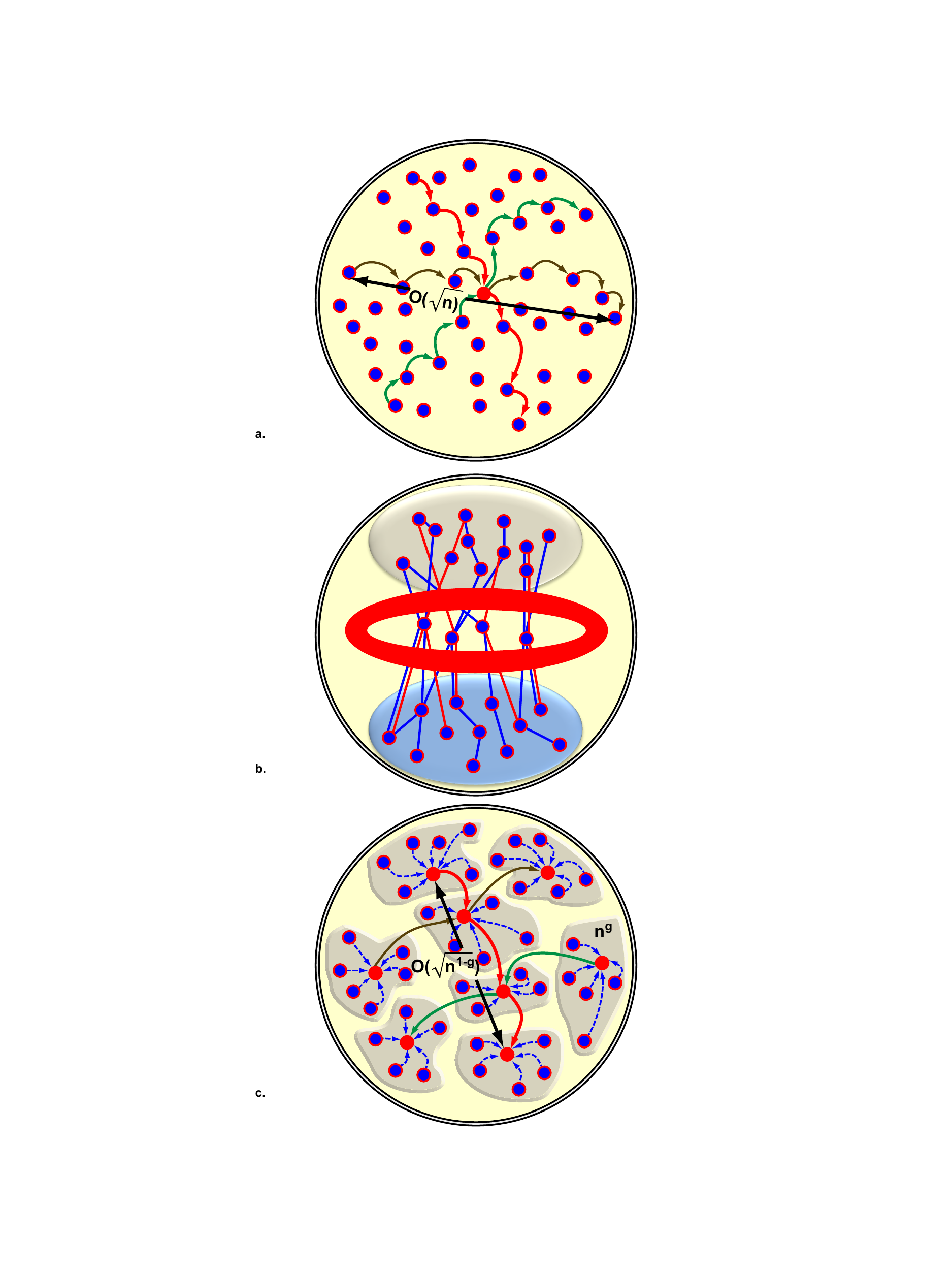}}
\caption{Bottlenecks, indicated by red nodes or enclosed in large red oval. {\bf a)} Multiple flows passing through an intermediate node must share the capacity of that intermediate node.  {\bf b)} Topological bottleneck {\bf c)} Cluster leaders form 
a coarse scaled version of the
the bottlenecks in a), when inter-cluster traffic must pass through these leaders.
}
\label{unicast_fig}
\end{figure}

We first consider a network which consists of $n$ nodes, each of which
has a message to send to another one of the $n$ nodes, all of which
lie in a fixed
two-dimensional circular area.  When $n$ is large, as $n$ increases, the diameter of the 
region  grows as $\sqrt{n}$. In order to minimize interference to other
nodes, each node relays its message across a path of  hops between
neighboring nodes towards its
destination, as illustrated in Figure~\ref{unicast_fig}a.  We consider the case that in order for most messages to reach their destinations, they must
traverse a number of hops   that scales as the diameter, or $\sqrt{n}$. Various deterministic and random node placements and
source-destination pair selection processes, such as a uniform distribution of nodes and node pairings, could yield this scaling
with the diameter.

Since each of the $n$ messages must be transmitted  by approximately
$\sqrt{n}$ relay nodes, a total of about $n \sqrt{n}$ relaying transmissions must be
made for the messages to reach their destinations.  The case with
minimum congestion would have these relaying transmissions distributed
uniformly across the $n$ nodes, so that each node performs about
$\sqrt{n}$  transmissions, one for each message it must relay.
Figure~\ref{unicast_fig}a illustrates  the bottleneck that forms at
one example relay node, as multiple messages pass through it.
 In the best case, each node can
relay one of the approximately
$\sqrt{n}$ messages in each unit of time.  More generally, a node may not be able to successfully relay a message to
its next hop in each time unit, owing to interference.  Therefore, each flow receives, at best,
a capacity that scales as  $1/\sqrt{n}$, as in \cite{GuptaKumar}.

In a three dimensional network most source-destination paths scale as
$n^{1/3}$ hops in length.  Hence the total number of relaying
transmissions for all $n$ messages would grow as $n^{4/3}$.  In the
best case, these transmissions are uniformly distributed among the
$n$ nodes, so that each node would need to relay about $n^{1/3}$
messages, yielding a per flow capacity that decreases at least as
fast as $n^{-1/3}$.
Therefore, capacity as a function of the number  of nodes in a fixed three dimensional region is increased relative to that in a fixed two dimensional region.  Henceforth, in this article we consider two-dimensional space.



The above analysis assumes the total link capacity does not change as the hop
length changes.  The same scaling behavior found here also holds if the
density of nodes is held fixed, and $n$ is instead increased by extending the
area of the network.


\subsection{When the Receivers are Bottlenecks: Multipoint Transmissions} \label{transmit}
We now discuss scaling  as a function of the number of transmitting nodes.   First, we consider the multipoint-to-point case, as is common when many nodes must communicate with a single base station or access point, 
for example.  It is assumed that  $n$ nodes each send a message to a single common destination node.  
Hence, the destination is a bottleneck at reception, which can allocate at
most $1/n$ of its capacity to each of the $n$ messages it is sent.

We next consider multipoint-to-multipoint connections in which $n$  transmitting nodes  each send a message to a common
multicast group {\it of any size} up to the broadcast size of $n$.  In
this case, each  node in the common multicast group needs to receive
$n$ messages.  Hence, each receiving node can allocate at most $1/n$ of its capacity to each transmitted message.  

These results can be extended to the case in which the $n$ transmitting nodes send messages to different multicast groups: The 
approximate $1/n$ capacity scaling holds for communication to any receiving node
that is a destination for some fixed fraction of the $n$  transmitting nodes.

In the cases discussed in this section, the {\it receiving nodes}
are the bottlenecks. These results hold regardless of mobility, topology, or reception capabilities.

\subsection{When the Network Topology Forms the Bottlenecks: All Connection Types}\label{bottle}

We consider networks which have the following common topology trait: somewhere in the network there is a bottleneck consisting of
 $B$ nodes, 
where $B$ is some constant independent of $n$,
and this bottleneck is located such that some  fraction $f$,
also independent of $n$, of the messages
must flow through it.
An example of such a topology is shown  in Figure~\ref{unicast_fig}b.
Each bottleneck node is thus shared by $fn$ flows.   Therefore, each
flow receives a fraction of the capacity that decreases as $1/n$.  This result does not depend on connection type,  mobility, nor  the precise topology.

\subsection{When the Clustering or Hierarchy   is the Bottleneck} \label{cluster}

Clustering and hierarchical structures are common in military networks, and questions arise as to whether these structures can improve scalability.   
We first consider communication among nodes, each of which is assigned to a cluster.  For the broadcast case, the argument focusing on the $n$ messages each receiving node must process, and the resulting scaling law, are the same as in Section~\ref{transmit}.  

We next turn to unicast traffic.  All inter-cluster communication must transit through cluster leaders, where a single leader is assigned  in  each cluster. 
We allow for the formation of
$n^{1-g}$ clusters, each of which  would consist of  $n^g$ nodes, as illustrated in Figure~\ref{unicast_fig}c, where $0 \leq g \leq 1$.

 We assume some fraction $f$, independent of $n$, of traffic is inter-cluster communication.
These $fn$ messages must be
relayed
through the  cluster leaders, which are indicated as red nodes in Figure~\ref{unicast_fig}c.
 Typical inter-cluster path lengths for these inter-cluster messages
 scale as the diameter, or about $n^{(1-g)/2}$ cluster
leader hops.

The situation for clustering is analogous to 
Figure~\ref{unicast_fig}a, except that  the flows that went through
each node in Figure~\ref{unicast_fig}a are now  constrained to
traverse only the cluster leaders in
Figure~\ref{unicast_fig}c. Therefore, the cluster leaders  are now the
bottlenecks.   The scaling argument for cluster leaders is analogous
to that in Section~\ref{fixed}:

The $fn$ inter-cluster messages must collectively be relayed a total
of $fn \times n^{(1-g)/2}$ hops.  In the best case, this relaying is
uniformly distributed among the $n^{1-g}$ cluster leaders, resulting
in each cluster leader needing to relay about $fn \times
n^{(1-g)/2}/n^{1-g} = fn^{(1/2 + g/2)}$ messages.  Therefore, in the
best case, the inter-cluster traffic is allocated a fraction of the
cluster leader's link capacity that scales as
$1/n^{(1/2 + g/2)}$.



This capacity 
is seen to be
maximized for $g = 0$, which corresponds to  clusters of size 1
node; in this case,  the  capacity obtained in the best case scales
as $n^{-1/2}$, which is the same as that discussed in Section~\ref{fixed}.
Forming clusters of a fixed size that is independent of $n$ will also yield the $g = 0$ result.
If larger clusters are formed by increasing $g$ for a given $n$,   the capacity is
{\it decreased}, according to the scaling law $1/n^{(1/2 + g/2)}$.
Forming clusters of size that increases with $n$ means fewer cluster leaders must carry the same total $fn$ flows, and hence each cluster leader would
then have
more flows among which it must divide its capacity.

The case $g = 1$ occurs when all $n$ nodes form a single cluster, in which case the scaling of capacity above reduces to $1/n$, which is the same scaling  found in Sections~\ref{transmit} and \ref{bottle}.  
In those sections,
the single cluster leader  corresponds to the base station or bottleneck node respectively.  Thus we have shown that the bottlenecks discussed in Sections~\ref{fixed} and \ref{transmit} plus \ref{bottle}  can be viewed as two extreme limiting cases of clustering for clusters of size 1 and $n$ respectively.

Given $n$, the length of hops between cluster leaders increases as
$g$ increases; hence,
in power limited networks, the {\it total} per link capacity decreases as $g$ increases. 
Therefore, in  a power limited network, the total capacity 
per inter-cluster message will actually decrease even more rapidly
with $g$ than $1/n^{(1/2 + g/2)}$.

 We now shift focus to  a multi-layer hierarchical topology which consists of $n$ nodes  
that communicate with a single base station in the hierarchy.
 Regardless of how the nodes are clustered in each layer, or how the inter-layer flows 
are regulated, capacity is still limited by the bottleneck of the  base station needing to receive $n$ flows, as described in Section~\ref{transmit}.
Therefore, the per flow capacity scales as $1/n$,  or worse, if the total per link capacity depends on $n$.

In summary, imposing clustering or hierarchical structure, does not
improve the scaling laws, and actually {\it decreases} the capacity 
in point-to-point communications when
cluster size grows with $n$, owing to the resulting bottlenecks formed.

\section{Methods to Avoid Bottlenecks} \label{no_bottleneck}

We explore the potential of  some methods that appeared recently in the literature to increase scalability.  In particular, we
consider whether these methods can  avoid the bottlenecks discussed in
Section~\ref{bottleneck} in practical networks.

\subsection{Mobility: Relaying with Mules} \label{mobility}

\begin{figure} [h!]
\centering
{\includegraphics[width=3.0in,keepaspectratio=1,clip=true,
viewport=1.75in 1.75in 8.25in 5in]{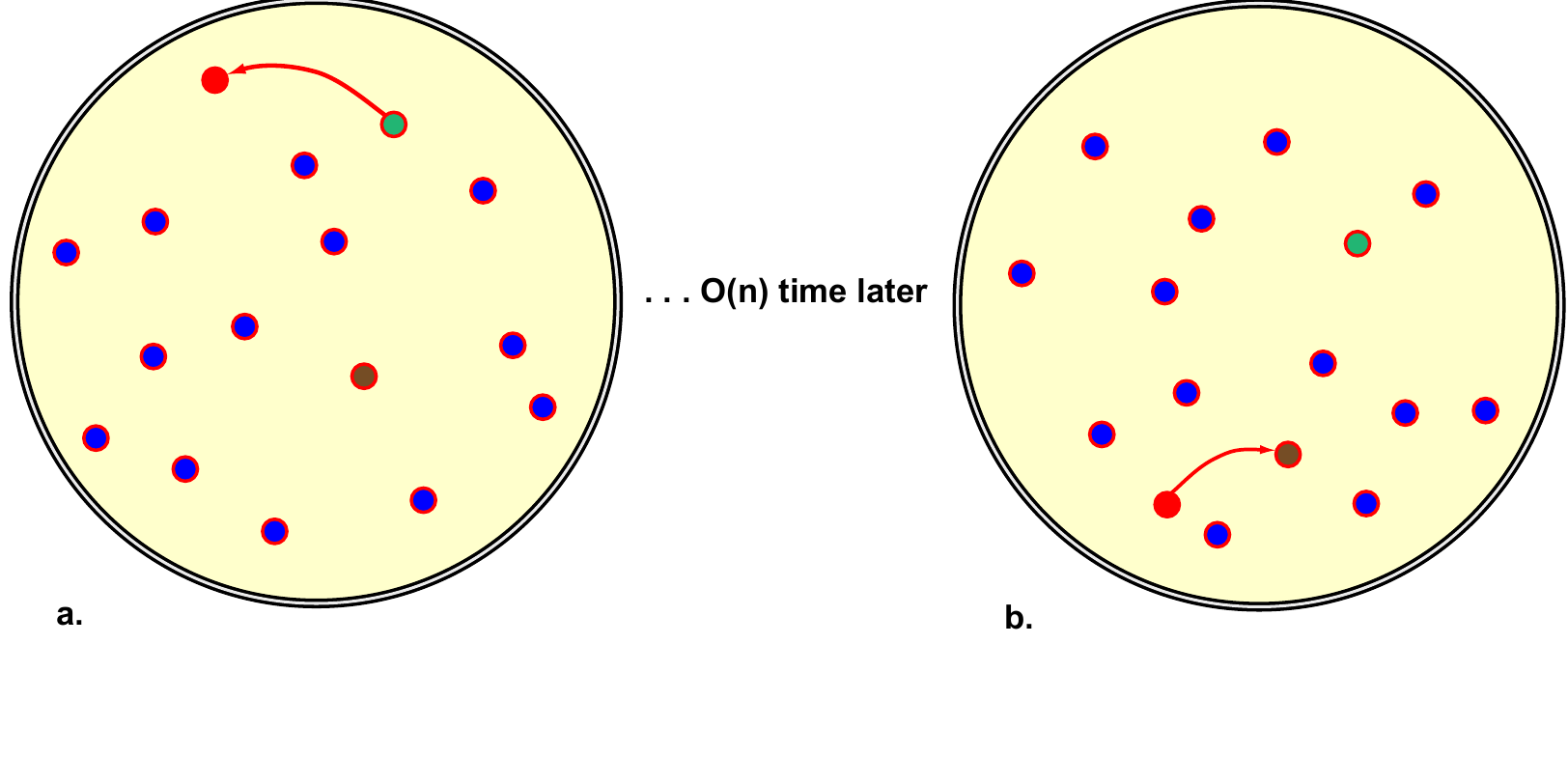}}
\caption{Relaying model of \cite{Grossglauser}: {\bf a)} A source
  node, colored green, sends its data to  another node, colored red, that comes within range.  This other node  then serves as one of a number of potential relay nodes for that source.
{\bf b)} At a point $O(n)$ later in time, a potential relay node, colored red, comes within range of a destination node.
}
\label{mobility_fig}
\end{figure}

First, we discuss mobility as a means to allow nodes to circumvent bottlenecks.
The work \cite{Grossglauser} considered the case  in which each node moves independently according to a stationary ergodic process.  In this model,  each node has a potential opportunity to transmit its packets to another nearby node at each unit in time; Figure~\ref{mobility_fig}a illustrates  transmission for one source node
to one  intermediate node.
{\it Each} such intermediate node  can henceforth serve as a potential relay node.   As the nodes move, there will be many such potential relay nodes for each source node, and hence it is likely that at least one of the potential relay nodes will {\it eventually} become close to the destination node.   Figure~\ref{mobility_fig}b illustrates  a later point in time when one relay node comes within range 
of a destination node. 

Each node thus sends and receives $O(1)$ traffic per unit time, as shown in \cite{Grossglauser}; thus capacity scales with network size.  However, the packets arrive at their destinations in a random order.   Moreover, the {\it delay} grows with the number of nodes, rendering such a scheme impractical in most large networks.

\subsection{Clustering \& Cooperation : Huge Virtual MIMO} \label{hier} 

Recent works in the literature, such as \cite{hier_Ozgur1} and \cite{hier_Niesen}, suggest that hierarchical networks
can improve scaling by cooperation.
In \cite{hier_Ozgur1}, for example, hierarchical cooperation is achieved by long distance MIMO transmissions across the network.  The sending and receiving MIMO arrays are virtual antenna arrays, each consisting of a cluster of a large number of nodes. 

We consider the implementability of these large scale virtual MIMO networks, as presented in \cite{hier_Ozgur1}.
First, note that  in order to achieve the MIMO capacities, the channel
must include multiple propagation paths. 
Most importantly, attaining the MIMO capacity is dependent on
receivers' knowledge of the channel.  There are $O(n^2)$ channel
conditions that most be transmitted periodically.    However, with the MIMO setup, there are only $O(n)$ parallel simultaneous communications that can occur.  Hence, if the channel update rate  is independent of $n$, the bandwidth required for these updates will grow  with $n$.
This issue is not addressed in \cite{hier_Ozgur1}, as static channels are assumed there.

\begin{figure} [h!]
\centering
{\includegraphics[width=3.5in,keepaspectratio=1,
viewport=1.75in 2.5in 8.25in 5.25in]{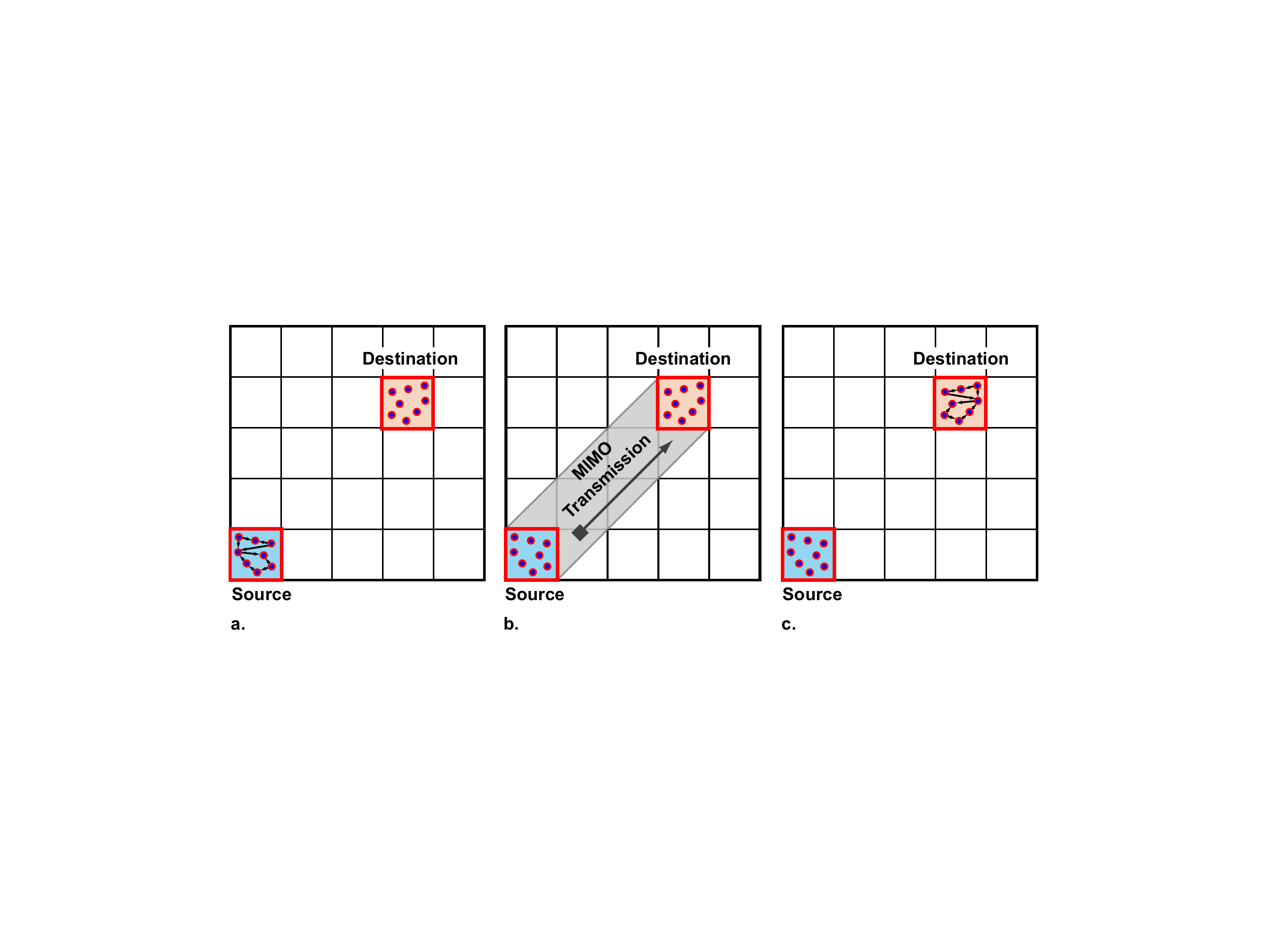}}
\caption{Three stages of the large virtual MIMO of \cite{hier_Ozgur1}  for a single representative source and destination cluster pair.
{\bf a)} Local exchange of information, simultaneously occurring in each source cluster. 
{\bf b)} One of many large scale MIMO transmissions.
{\bf c)} Local exchange of quantized information simultaneously occurring within each destination cluster to perform decoding.
}
\label{MIMO_fig}
\end{figure}

We next consider additional assumptions of \cite{hier_Ozgur1}. Before a long distance MIMO transmission commences, each node must distribute data to all other nodes in its cluster. This stage is illustrated for one  source cluster in Figure~\ref{MIMO_fig}a.  It is assumed \cite{hier_Ozgur1} that many source clusters are performing these initial distributions in parallel, based on the assumption \cite{hier_Ozgur1} of spectrum segregation among clusters.  We note that as node density increases, the assumption of spectrum segregation of clusters that are within range of one another requires a corresponding increase in bandwidth;  at  a high enough node density, the required bandwidth can become untenable.

After the long distance MIMO transmission, which is illustrated in
Figure~\ref{MIMO_fig}b, the nodes cooperate to decode, as depicted in
Figure~\ref{MIMO_fig}c. The assumption \cite{hier_Ozgur1} of scale
invariance is used twice in this decoding stage: First, it is assumed
that the combined signal each node receives for each destination node
in its cluster 
can be quantized to a number of bits that is independent  of $n$. We note that when $n$ is increased by extending the area of the network, thereby extending the distances of the MIMO transmissions in Figure~\ref{MIMO_fig}b, the SNR received at each node
in the receive clusters decreases. In this case, quantization with
insufficient granularity would cause distortion.
Secondly, even with sufficiently fine scaled quantization, the resulting low SNR from MIMO transmission of Figure~\ref{MIMO_fig}b in extended networks can itself hamper the decoding of Figure~\ref{MIMO_fig}c.

The implications described above of the spectrum segregation and scale
invariance assumptions in the dense and extended network types are
summarized in Table 1. It is seen that as the number of nodes becomes
large, each of these assumptions complicates implementation of a large
scale MIMO scheme for a different network type.  
Hence, in many practical scenarios  without infinite bandwidth or
infinite power, scaling laws with clustering would instead follow those derived in Section~\ref{cluster}.

\begin{table}[h!] 
\scriptsize
\begin{tabular}{|c||c|c|c|}
	\hline 
Assumption & Dense & Extended & Stage Required \\
\hline \hline
Quantization \&  & & & \\
SNR Scale invariance  & $\surd$ &  X & c\\
\hline
Spectrum Segmentation & X & $\surd$ & a, c\\
\hline
\end{tabular}
\caption{Indication of whether each assumption used in a large  hierarchical MIMO scheme, such as in \cite{hier_Ozgur1}, can be implemented without very large power or bandwidth in dense or extended networks, as the number of nodes grows very large.}
\end{table}

\section{Erasures} \label{erasures}
The issue of low SNR discussed in the last section leads more
generally to the question of the impact of  erasures on scalability. 
This question was not addressed in earlier sections here, nor in much
of the literature. 
We now discuss how various erasure recovery protocols  affect  scalability.
Furthermore, we illustrate the cost of
adding more receiving nodes to multicast transmissions, when erasures are considered.

\subsection{End-to-End Retransmissions: Point-to-Point} \label{tcp}

\begin{figure} [h!]
\centering
{\includegraphics[width=2.5in,keepaspectratio=1,
viewport=2.5in 3.75in 5in 6.25in]{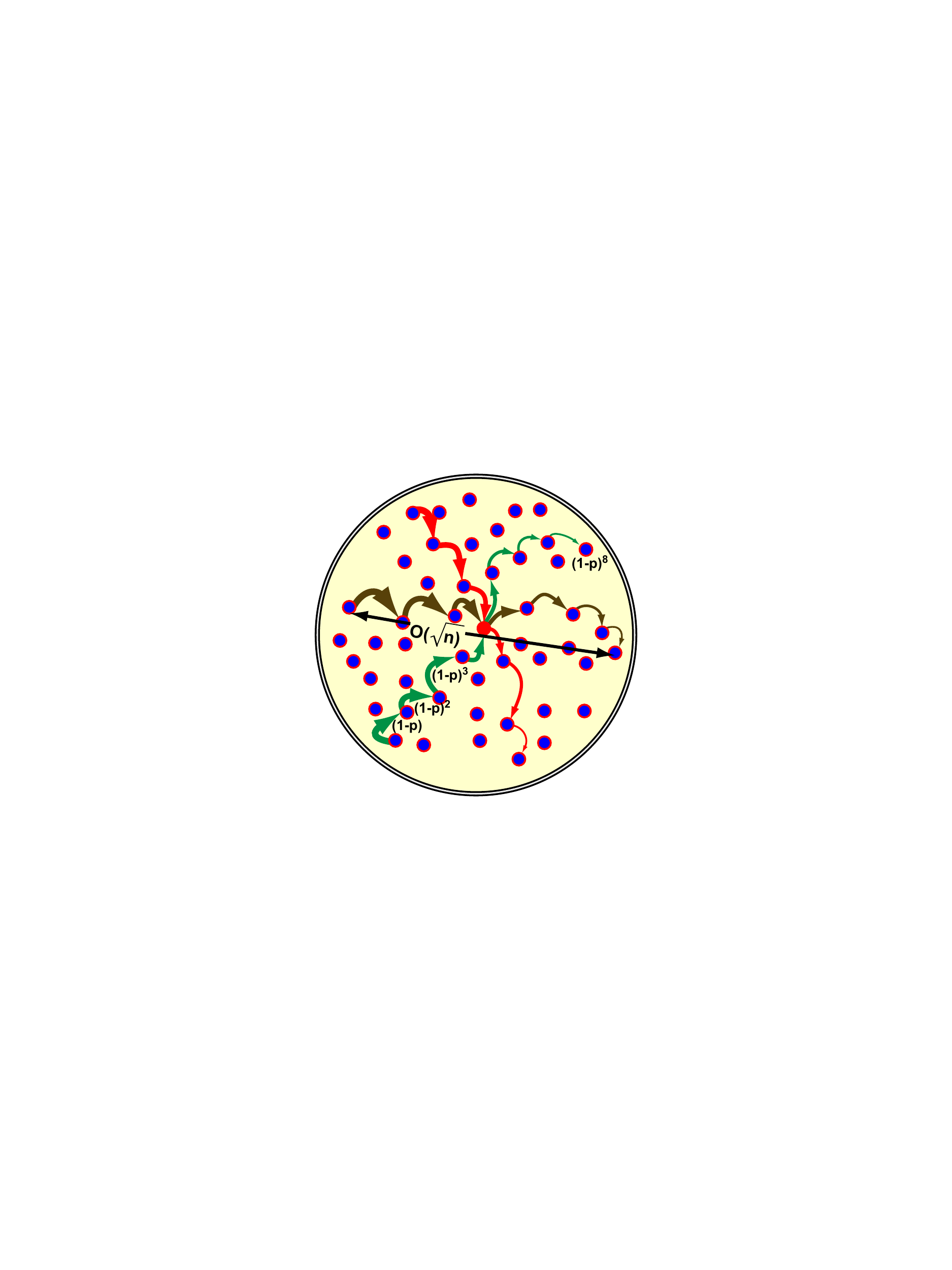}}
\caption{Probability of successful transmission across a path: decreasing arrow thickness with each hop represents  the decrease in probability that a message arrives at the node at the end of the corresponding hop.
}
\label{endtoend}
\end{figure}

End-to-end acknowledgements are used in protocols such as TCP.
If  the erasure probability  on each link is $p$, then the probability a packet reaches its destination at the end of the path is given by $(1-p)^{C\sqrt{n}}$, since a packet must not be lost at each of the $C\sqrt{n}$ hops, where $C$ is a constant independent of $n$. The decrease in probability of a message reaching successively distant hops is illustrated in Figure~\ref{endtoend}.

Therefore, since
$\lim_{n \to \infty} (1-p)^{C\sqrt{n}} = 0$,
each packet's probability of reaching its destination approaches zero!  The throughput per flow decreases exponentially as  $(1/\sqrt{n}) \times (1-p)^{C\sqrt{n}}$.  Thus in this case, the scaling is dominated by the erasures only being corrected on an end-to-end basis, and the throughput decreases much more rapidly than the model used in Section~\ref{fixed} with no erasures.

We have assumed $p$ is independent of $n$, which is always the case in an extended network, as more nodes are added in that case by increasing the area.
This assumption also holds in a dense network, if the decrease in the nodes' transmit powers necessitated to avoid the interference of 
the added nodes is matched to yield the original erasure probability.

\subsection{Link Layer Retransmissions: Point-to-Point} \label{link_layer}
We now consider the case in which  acknowledgments are sent on a hop-by-hop basis.
Assuming an erasure probability $p$ on each link, 
the throughput is reduced by $1-p$; hence, it   scales as $1/\sqrt{n} \times (1-p) \sim 1/\sqrt{n}$. This result obeys the same scaling law as the case of no erasures.

\subsection{Single Hop Multicast} \label{receive} 
Scaling of multicast message delivery time with the number of receiving nodes can be dominated by the recovery of erasures.
For best effort delivery, there is no penalty to adding any number of receiving nodes, as there is no erasure recovery.

In contrast, for reliable delivery, additional packets must be sent
when there are erasures, until {\it every node}  receives all packets.  
We consider a multicast group of $n$ nodes that must receive a file from a single transmitting node.  

In \cite{arxiv} an efficient method  using a  predictive model and coding are used for multicast delivery, so that
erasure recovery is accomplished by the preemptive transmission of the expected number of coded packets needed to complete transmission to all recipients with high probability.   Any given coded packet can be used by different recipients to recover from different erased packets. With this method, the total completion time for all nodes to receive a file depends only very weakly on $n$ \cite{arxiv}.   For example, in order to obtain a $90\%$ probability that $n= 10$ users all receive a file of length 100 time slots, when the channel's packet erasure probability is $10\%$, a total of $125$ time slots would be required \cite{arxiv}.  If the same file were sent over the same channel, but the number of recipients were increased from $10$ to $1000$, then the total time needed to complete delivery to all nodes with 
$90\%$ probability would increase to only $142$ time slots \cite{arxiv}.
  It is further shown \cite{arxiv} that for all practical values  of $n$, the dependence on file size, rather than  dependence on the number of users, dominates the behavior of completion time.  Hence, it is important when focusing on scaling with the number of users to also consider if other factors may actually contribute more to the metric of interest.  

The average per packet completion time for reception by all users decreases with increasing file size with efficient multicast transmissions \cite{arxiv}.
Future research could leverage this favorable  scaling with file size to artificially construct larger files for more efficient channel use. 
For example, we now depart from the case of a single source to consider the case in which $J$ transmitting nodes have files to send to
a multicast group.
If  transmitting node $i$ has a block of  size $k_i$, a total of $J$ transmitting nodes can collectively create a  block of size $ k = \sum_{i=1}^J k_i$.  Each of the $J-1$ nodes that transmits after the first node transmits would code together the packets it receives from the preceding transmitting nodes with its own source packets so as to synthesize a larger block of size $k$.  
The cost of using this method to decrease the mean per packet completion time to the entire group is an increase in delay for the some of the $n$
nodes to completely receive the blocks from the earlier transmitting
nodes.  

This synthesis of a large block size through collaboration has the additional advantage of allowing the first transmitting nodes to cease retransmissions, once another transmitting node has received their information.  Furthermore, it can assist in an extended network so that closer nodes can serve to distribute the data of distant nodes, and can thereby also conserve battery power. Detailed exploration of the benefits of such a scheme are left as future work.

\section{Summary and Discussion} \label{summary}
We have provided a simple overview of some representative studies on scaling.  We have  emphasized the import of bottlenecks and erasures,  to show that scaling laws depend significantly  on the protocols and assumptions used.  We have illustrated how  some of the  scaling laws derived in the literature will be difficult to implement in practice, owing to their idealized assumptions.  Furthermore, we have shown that  under a range of practical scenarios that include multicast traffic,
topological bottlenecks, and hierarchical structures, the per node
capacity scales as  $1/n$.

A number of additional scaling issues not discussed here
 are a subject for future investigation.  While this article analyzed
scalability in terms of the data, there
still remains
the issue of scalability of 
network control messages.  
In addition, we note that the bottlenecks discussed here
also contribute to a potential scaling problem of storage or memory at
the bottleneck nodes, as well as associated queueing delays.  

\end{document}